\def\tr{{\text{tr}}\,}
\def\Tr{{\text{Tr}}\,}

\def\be{\begin{equation}}
\def\ee{\end{equation}}
\def\bea{\begin{eqnarray}}
\def\eea{\end{eqnarray}}
\def\bse{\begin{subequations}}
\def\ese{\end{subequations}}

\documentclass[prl,twocolumn,showpacs,amsmath,amssymb]{revtex4}
\usepackage{graphicx}
\usepackage{dcolumn}
\usepackage{bm}
\begin{document}
\title{Signatures of pairing mechanisms and order parameters in ferromagnetic 
                                                       superconductors\\
}
\author{T.R. Kirkpatrick}
\affiliation{Institute for Physical Science and Technology, and Department of
         Physics\\
         University of Maryland, College Park, MD 20742}
\author{D.Belitz}
\affiliation{Department of Physics and Materials Science Institute, 
         University of Oregon, 
         Eugene, OR 97403}
\date{\today}

\begin{abstract}
Two predictions are made for properties of the ferromagnetic 
superconductors discovered recently. The first one is that
spin-triplet, p-wave pairing in such materials will give the
magnons a mass inversely proportional to the square of the 
magnetization. The second one is based on a specific mechanism 
for p-wave pairing, and predicts that the observed broad anomaly 
in the specific heat of URhGe will be resolved into a split 
transition with increasing sample quality. These predictions 
will help discriminate between different possible mechanisms 
for ferromagnetic superconductivity.  
\end{abstract}


\maketitle

Recently there has been a considerable amount of work on what is believed 
to be true coexistence between ferromagnetism and superconductivity. In 
UGe$_2$ \cite{Saxena_et_al_00}, ZrZn$_2$ \cite{Pfleiderer_et_al_01},  and 
URhGe \cite{Aoki_et_al_01}, superconductivity has been observed at
very low temperatures inside the ferromagnetic phase. 

There have been several theoretical suggestions for the cause of this 
exotic superconductivity. Early theories proposed magnetically mediated
spin-triplet superconductivity in ZrZn$_2$ \cite{Fay_Appel_80}, and
recent band-structure calculations have concluded that this is a possibility
in these materials \cite{Shick_Picket_01, Santi_et_al_01}, 
and Machida and Ohmi \cite{Machida_Ohmi_01} have give arguments for the 
pairing to be of p-wave, non-unitary spin-triplet nature. On the other hand, 
a conventional phonon mechanism has been proposed in 
Ref.\ \onlinecite{Shimahara_Kohmoto_02}, and 
Ref.\ \onlinecite{Watanabe_Miyake_02} has suggested that the superconductivity 
is due to coupled CDW and SDW fluctuations. It is not obvious from these 
considerations why the superconductivity is observed {\em only} in the 
ferromagnetic phase. To explain this,
Sandeman et al. have proposed a density-of-states effect, that exists 
only in the ferromagnetic phase, as the source of the superconductivity 
in UGe$_2$ \cite{Sandeman_et_al_03}. Kirkpatrick et. al.\ 
\cite{us_p-wave_letter,us_p-wave} have proposed an explanation for the
observed phase diagram that is based on an enhancement of the longitudinal
spin susceptibility in the ferromagnetic phase by magnons, or magnetic
Goldstone modes. 

In this Letter we predict two observable effects that allow to partially
discriminate between these various theoretical mechanisms. First, we
show that if the order parameter (OP) is of non-unitary spin-triplet type,
the ferromagnetic magnons, or transverse spin fluctuations, become 
effectively massive
in the superconducting phase, with a mass that is inversely proportional 
to the square of the magnetization. This prediction depends only on the
symmetry of the OP, and is independent of the mechanism for 
superconductivity. The second prediction is based on our previously 
suggested mechanism for superconductivity in these systems 
\cite{us_p-wave_letter,us_p-wave}.
It predicts that the specific heat anomaly observed in URhGe
around the superconducting transition temperature \cite{Aoki_et_al_01}
actually consist of two distinct features that will be resolved as the
sample quality increases. These two features correspond to the pairing
of electron spins parallel and antiparallel, respectively, to the direction
of the magnetization. In what follows, we will first give our detailed
results and simple intuitive explanations, and then we will sketch their
technical derivation \cite{us_unpublished_03}.

General symmetry principles \cite{Forster_75} show that, in the presence 
of a nonvanishing magnetization in $z$-direction, the magnetic
susceptibility tensor has the structure 
\be
\chi = \begin{pmatrix} \chi_{\text{T},+} & \chi_{\text{T},-} & 0 \\
                      -\chi_{\text{T},-} & \chi_{\text{T},+} & 0 \\
                       0 & 0 & \chi_{\text{L}}
       \end{pmatrix}\quad,
\label{eq:1}
\ee
with $\chi_{\text{T},\pm}$ and $\chi_{\text{L}}$ the transverse and 
longitudinal magnetic susceptibilities, respectively. In the absence of
superconductivity, $\chi_{\text{T}}$ is massless, since it is the 
magnetic Goldstone mode \cite{Forster_75}. Spin-triplet superconductivity 
explicitly breaks the spin rotation symmetry and leads to a mass in 
$\chi_{\text{T}}$ \cite{Goldstone_mode_footnote}.
For small frequencies $\Omega$ and wavevectors $\bm{k}$ we find
\be
\chi_{\text{T},\pm}(\bm{k},i\Omega) = 
   \left[\frac{a_{\pm}\,\tilde{\delta}} {i\tilde{\Omega} + \tilde{\delta}\,
      b\,(\tilde{\bm{k}}^2 + \mu^2)} \pm (\delta \rightarrow -\delta)
             \right]\quad.
\label{eq:2}
\ee
Here $\tilde{\Omega} = \Omega/4\epsilon_{\text{F}}$ and 
$\tilde{\bm{k}} = \bm{k}/2k_{\text{F}}$ are the frequency and wavevector 
made dimensionless by means of the Fermi energy $\epsilon_{\text{F}}$ and 
the Fermi wavenumber $k_{\text{F}}$, respectively. $a_+$ and $b$ are real 
constants on the order of unity, and $a_-$ is imaginary with a modulus on 
the order of unity. $\tilde{\delta} = \delta/4\epsilon_{\text{F}}$, with 
$\delta$ the Stoner splitting of the Fermi surface, which is proportional 
to the magnetization $m$. In a phase where only up-spin electrons are paired,
with $\Delta_{\uparrow}$ the up-spin superconducting energy gap, we find 
for the mass
\bse
\label{eqs:3}
\be
\mu^2 = (\Delta_{\uparrow}/2\delta)^2\,f(\delta/2T)\quad,
\label{eq:3a}
\ee
where
\be
f(x) = \int_0^{\infty}\frac{dy}{y(1-y^2)}\ \tanh(yx)\quad.
\label{eq:3b}
\ee
\ese
Asymptotic analysis yields $f(x\rightarrow\infty) = \ln x$, and
$f(x\rightarrow 0) = 0.85\ldots\times x^2$. Such a mass in the transverse 
magnetic susceptibility is in principle observable by neutron scattering.
An estimate using $\Delta_{\uparrow} \approx 2T_{\text{c}\uparrow}
= 1\,{\text{K}}$, $\delta \approx \epsilon_{\text{F}}/2 = 10^4\,{\text{K}}$, 
and $k_{\text{F}} = 1\,{\AA}^{-1}$,
yields $2k_{\text{F}}\mu = 3\times 10^{-4}\,{\AA}^{-1}$. This is roughly a 
factor of $30$ smaller than the current lower wavenumber limit for 
inelastic neutron scattering \cite{Aronson_03}. Since $\mu$ depends roughly 
linearly on $\Delta_{\uparrow}$, a higher value of $T_{\text{c}}$ would
help push $\mu$ into an observable regime.  

\begin{figure}[t]
\includegraphics[width=7cm]{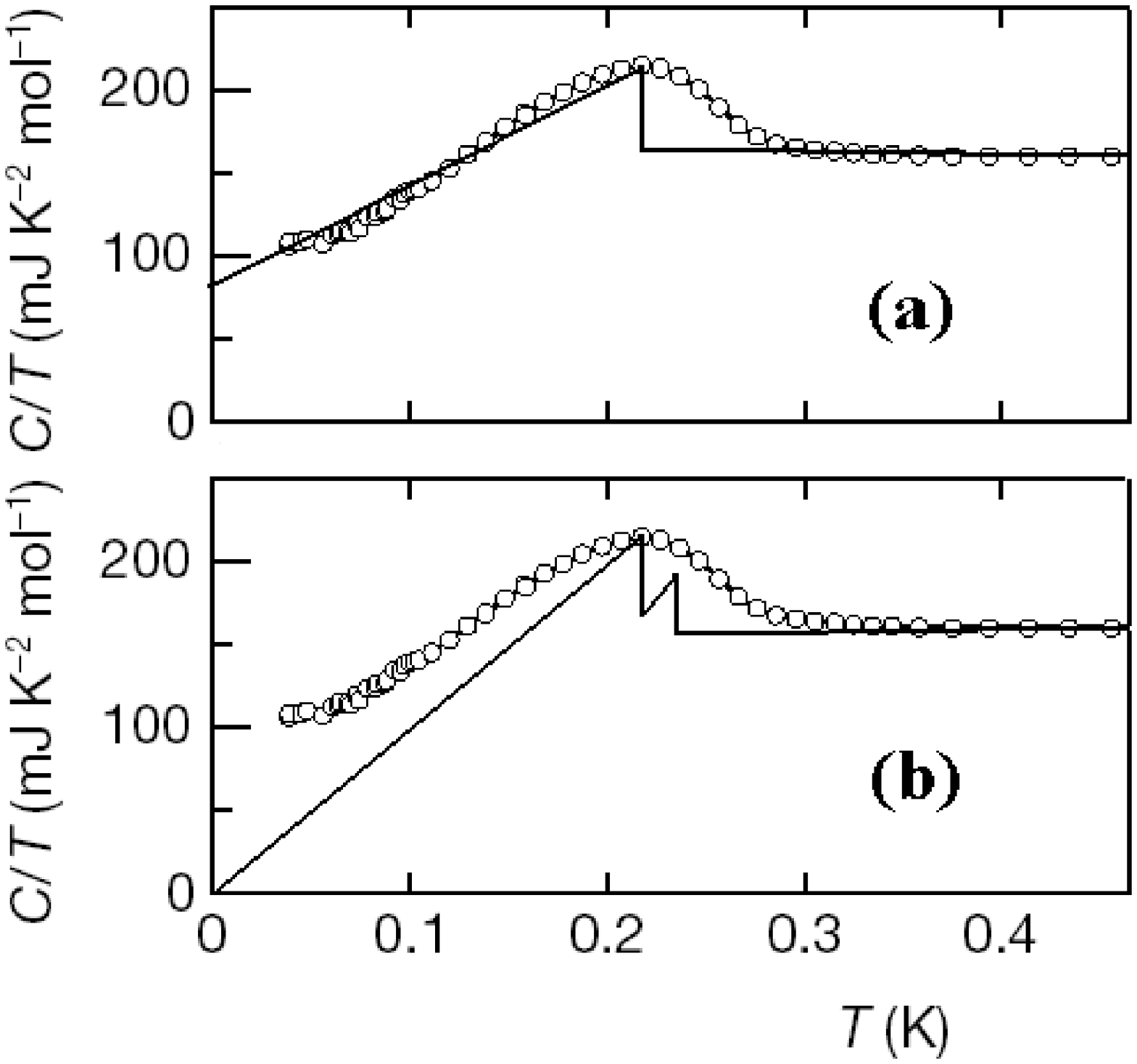}
\vskip -30pt
\caption{\label{fig:1} Circles denote the experimental specific heat 
 coefficient of URhGe \protect{\cite{Aoki_et_al_01}}. The
 solid lines show the theoretical expectation if only half of the
 electrons pair (a), and if there are two transitions within 5\% of
 one another (b). See the text for additional information.}
\vskip -5mm
\end{figure}

Our second prediction concerns the detailed structure of the specific
heat anomaly oberserved in URhGe \cite{Aoki_et_al_01}. The experiment
shows a broad peak with a maximum close to the resistive $T_c$,
which is evidence for true bulk superconductivity. However, for
low temperatures the specific heat coefficient does not go to zero, 
but rather approaches a finite value that is roughly half its value
in the normal phase, see Fig. \ref{fig:1}. Ref.\ \onlinecite{Aoki_et_al_01}
has interpreted this behavior as an indication for the pairing being
restricted to one spin projection, while the electrons with the opposite
spin projection retain a Fermi surface down to the lowest temperatures
measured. With a relative specific heat discontinuity $\Delta C/C$ adjusted
to fit the data, and a temperature dependence below $T_c$ given by 
$C(T)/T \propto T$ \cite{specific_heat_footnote},
this suggests that, with improving sample quality, the data should
approach the solid curve shown in Fig.\ \ref{fig:1}(a).
This is not the only possible interpretation. We have extended our 
previous calculation of the superconducting
$T_c$ to include both up-spin and down-spin pairing. For the superconducting
gap, which is a $2\times 2$ matrix $\hat{\Delta}$ in spin space, we assume
\cite{orbital_structure_footnote}
\be
{\hat\Delta}({\bm k}) = \begin{pmatrix} 
      \Delta_{\uparrow}\hat{k}_z & 0 \\
      0 & \Delta_{\downarrow}\hat{k}_z
         \end{pmatrix}\quad,
\label{eq:4}
\ee
with $\hat{\bm{k}}$ a unit wavevector. This is the OP of the 
A$_2$ phase in He$^3$ \cite{Vollhardt_Woelfle_90}.
For generic parameter values we find that there is only a small difference
between the values of $T_{\text{c}\uparrow}$ and $T_{\text{c}\downarrow}$ in 
the ferromagnetic phase, on the order of 5-10\%. This is comparable to
the difference Fay and Appel \cite{Fay_Appel_80} found within a simpler
theory that could not describe the strong enhancement of $T_c$ in the
ferromagnetic phase. Consequently, one expects that in a perfect
sample the specific heat data would qualitative look as shown in
Fig.\ \ref{fig:1}(b). In this interpretation, the broad peak hides
a split transition, and the low-temperature tail of the specific
heat coefficient is due to non-superconducting parts of the sample.
(See, however, Ref.\ \onlinecite{finite_gamma_footnote}.)
It is interesting that such a split transition has been
observed in the heavy-fermion superconductor UPt$_3$ 
\cite{Fisher_et_al_89,Hasselbach_et_al_89} (which is {\em not}
ferromagnetic, and where the splitting has a different physical origin),
but only after a long period of improving sample qualities.

We now turn to the origin of these results, starting
with the mass in the magnon. Any spin-triplet OP can be
represented as a vector $\bm{\phi}$ in spin space that is isomorphic to
the $2\times 2$ spin matrix representation \cite{Vollhardt_Woelfle_90}.
For the OP given in Eq.\ (\ref{eq:4}), this vector is 
proportional to 
\[ \bm{\phi} = (\Delta_{\uparrow} - \Delta_{\downarrow},
     i(\Delta_{\uparrow} + \Delta_{\downarrow}),0)\quad. \]
We see that $\bm{\phi}^*$
is not parallel to $\bm{\phi}$. This an example of a ``non-unitary'' OP,
and Ref. \onlinecite{Machida_Ohmi_01} has argued that any
OP in ferromagnetic superconductors is likely of this general
type. In a Landau theory, the lowest order term in the free energy density
that couples the magnetization vector $\bm{M} = (0,0,m)$ and the 
superconducting OP that is allowed by symmetry is thus of the 
form
\be
f_{\text{coupling}} = ic\bm{M}\cdot\left(\bm{\phi}\times\bm{\phi}^*\right)
                        \quad,
\label{eq:5}
\ee
with $c$ a real constant. This shows that the presence of a nonzero 
superconducting OP, $\Delta_{\uparrow}$ is equivalent to an 
effective magnetic field $h_{\text{eff}} \propto \Delta_{\uparrow}^2$.
General symmetry considerations \cite{Ma_76} 
show that the transverse magnetic susceptibility must then have a mass
that is given by $\chi_{\text{T}}^{-1}(0,0) = h_{\text{eff}}/m$.
This explains the existence of a mass, and the dependence of $\mu^2$, 
Eq.\ (\ref{eq:3a}), on $\Delta_{\uparrow}$. 

For an understanding of the remaining structure of $\mu^2$, as well as
of our second prediction, we return to our previous work on the
superconducting transition of the up-spin electrons \cite{us_p-wave}.
As in that paper, we consider a model of free electrons with a static,
point-like spin-triplet interaction with amplitude $\Gamma_{\text{t}}$.
For superconducting OP fields we choose, 
$\mathcal{F}_{\uparrow}(x,y) = \psi_{\uparrow}(x)\psi_{\uparrow}(y)$ and 
$\mathcal{F}_{\downarrow}(x,y) = \psi_{\downarrow}(x)\psi_{\downarrow }(y)$,
with $\psi_{\sigma }(x)$ an electronic field with spin index $\sigma$
and space-time index $x$. The OPs, i.e., the expectation values 
$\langle\mathcal{F}_{\sigma}(x,y)\rangle = F_{\sigma}(x-y)$, are the 
anomalous Green functions. 

Using a technique similar to that employed in Ref.\ \onlinecite{us_p-wave},
we have derived coupled equations of motion for the $F_{\sigma}$, the
normal Green functions $G_{\sigma}$, and the magnetization $m$, in a
systematic loop expansion. Because our pairing mechanism is
due to magnetic fluctuations, one needs to go to one-loop order in order
to obtain superconductivity. To that order, and for $p$-wave orbital
symmetry, we find the following
Eliashberg-type equations for the gap functions $\Delta_{\sigma}$ and
the normal self-energies $\Sigma_{\sigma}$,
\bse
\label{eqs:6}
\bea
\Delta_{\sigma}(k)&=&\Gamma _{t}\int_{q}\chi _{L}(k-q)
                 \Delta _{\sigma}(q)/d_{\sigma }(q)\quad,
\label{eq:6a}\\
\Sigma_{\sigma }(k)&=&\Gamma_{\text{t}}\int_{q}\chi_{\text{L}}(k-q)
   G_{\sigma}^{-1}(q)/d_{\sigma }(q) 
\nonumber\\
&&\hskip -50pt + 2\Gamma_{\text{t}} \int_{q}[\chi_{\text{T},+}(k-q)
         +i\chi_{\text{T},-}(k-q)]\,G_{-\sigma}^{-1}(q)/d_{-\sigma}(q)\ , 
\nonumber\\
\label{eq:6b}
\eea
\ese%
with $\sigma = \uparrow,\downarrow \equiv +,-$. Here we use a 
wavevector-frequency four-vector notation $k=(\mathbf{k},i\omega_n)$, and 
$\int_q=T\sum_{n}(1/V)\sum_{\mathbf{q}}$. We
have introduced $d_{\sigma}(k) = G_{\sigma }^{-1}(k)\,G_{\sigma}^{-1}(-k)
+\Delta_{\sigma }(k)\Delta_{\sigma}^{\dagger}(k)$, and the
$G_{\sigma}^{-1}(k) = i\omega_n - (\bm{k}^{2}/2m_{e} - \mu - \sigma\delta)
 - \Sigma_{\sigma}(k)$ are the inverse `normal' Green functions, where 
$\delta =\Gamma_{\text{t}} m$ is the Stoner energy. The magnetic equation 
of state to one-loop order is given by,
\bea
m&=&\frac{T}{2V}\Tr(\gamma_3\mathcal{G}) + \frac{\Gamma_{\text{t}}}{2}
   \int dx\,dy \sum_{jk} \chi_{jk}(x-y)\,
\nonumber\\
&&\times \tr\left[\gamma_3\,\mathcal{G}(-x)\,\gamma_j\,\mathcal{G}(x-y)\,
   \gamma_k\,\mathcal{G}(y)\right]\quad.  
\label{eq:7}
\eea

It is easy to show that the magnetic susceptibility $\chi$ that serves
as the effective potential in these equations is proportional to the
physical spin susceptibility, Eq.\ (\ref{eq:1}). In a Gaussian
approximation \cite{RPA_footnote} we find
\be
\chi_{ij}^{-1}(x-y) = \delta_{ij}\,\delta(x-y) + \frac{\Gamma_{\text{t}}}{2}
   \,\tr\left[\mathcal{G}(y-x)\,\gamma_i\,\mathcal{G}(x-y)\,\gamma_j\,\right]
    \,.
\label{eq:8}
\ee
In these equations, the $(\gamma_0,\bm{\gamma}) = (\sigma_3\otimes
\sigma_0,\sigma_3\otimes\sigma_1,\sigma_0\otimes\sigma_2,\sigma_3\otimes
\sigma_3)$ are generalized Nambu matrices, and the $\mathcal{G}$ are
$4\times 4$-matrix Green functions that represent the normal 
and anomalous Green functions in this
basis. $\Tr$ denotes a trace over all degrees of freedom, and $\tr$ denotes a
trace over the discrete degrees of freedom.

An evaluation of Eq.\ (\ref{eq:8}) leads to a transverse susceptibility
tensor that has the form of Eq.\ (\ref{eq:2}), with $a_+ = 1$,
$a_- = -i$, $b = 2/3$, and $\mu^2$ as given in Eq.\ (\ref{eq:3a}).
The general Landau theory argument given above, see Eq.\ (\ref{eq:5}),
shows that this form of $\chi$ is generic, and not an artifact
of our Gaussian approximation. (See, however, 
Ref. \onlinecite{Goldstone_mode_footnote}.) The explicit calculation also shows 
that the prefactor of the $\Delta_{\uparrow}^2$ in the mass is a singular
function of the magnetization and the temperature. This implies that
a strict Landau expansion in powers of both the magnetic and the
superconducting OPs is singular, since the coefficients
of the Landau function do not exist. This breakdown of the Landau
expansion is due to the soft particle-hole excitations in an itinerant
electron system, and is analogous to an effect that has been discussed
for quantum ferromagnets in the absence of superconductivity
\cite{us_magnon_dispersion}.

Our second prediction hinges on a solution of the gap equations,
Eqs.\ (\ref{eqs:6}), which require the magnetic susceptibility tensor
$\chi$ as an input. A complete numerical solution of these strong-coupling 
equations would very difficult, and
does not seem worthwhile in the absence of detailed experimental information
about $\chi$. In Ref.\ \onlinecite{us_p-wave} we emphasized that
the relative values of $T_{\text{c},\uparrow}$ in the paramagnetic and
ferromagnetic phases, respectively, can be obtained within a simple
McMillan-type approximation, even though the absolute values obtained
by this method are probably not reliable. Here we adopt the same philosophy,
with the aim being to compute the relative magnitudes of both 
$T_{\text{c},\uparrow}$ and $T_{\text{c}\downarrow}$. 
In the paramagnetic phase this is straightforward, using the Gaussian
approximation, Eq.\ (\ref{eq:8}), for the magnetic susceptibility.
We find results for $T_{\text{c},\uparrow}$ and $T_{\text{c}\downarrow}$
that are comparable to those obtained by Fay and Appel \cite{Fay_Appel_80},
i.e., a ratio $T_{\text{c},\uparrow}/T_{\text{c}\downarrow}$ that is,
for generic parameter values, on the order of $1.1$.

In the ferromagnetic phase we need to take into account the effect
discussed in detail in Ref.\ \onlinecite{us_p-wave}, namely an enhancement
of the longitudinal susceptibility due to a coupling of the latter to the
ferromagnetic Goldstone modes \cite{Brezin_Wallace_73}. This requires
a calculation of $\chi$ to one-loop order, beyond the Gaussian
approximation given in Eq.\ (\ref{eq:8}). For $T_{\text{c}\uparrow}$ 
this calculation has been performed in Ref.\ \onlinecite{us_p-wave}.
For $T_{\text{c}\downarrow}$ there is the additional complication
that inside the superconducting phase, $\chi$ depends on the up-spin gap 
$\Delta_{\uparrow}$. This introduces a feedback effect that is
characteristic of any purely electronic mechanism for superconductivity.
Since the overall effect of this feedback is rather
small (see below), we can approximate the temperature
dependent $\Delta_{\uparrow}$ in $\chi$ by $2T_{\text{c}\uparrow}$.
We have found that $T_{\text{c}\downarrow}$, like its
$T_{\text{c}\uparrow}$ counterpart, is enhanced over its value in the
paramagnetic phase by a factor of 50-100, and for generic parameter
values it is within about 10\% of $T_{\text{c}\uparrow}$.
For the specific heat this leads to the prediction shown qualitatively
in Fig.\ \ref{fig:1}, namely, two transitions in close proximity.
Our calculations therefore suggest that the specific heat peak measured 
in URhGe actually consist of two unresolved peaks. We note that the 
width of the experimentally observed peak easily encompasses
both mean-field discontinuities. In contrast to our
first prediction, the second one is specific to the pairing mechanism
we have considered. If experiments on cleaner samples should show no 
indications of two closely spaced transitions, that would be
a strong argument against the spin-fluctuation induced pairing we
have considered.

We finally discuss our second result in some
more detail. The salient point of our mechanism for a strongly
enhanced superconducing $T_{\text{c}}$ in the ferromagnetic phase is the
coupling of the massless (on the superconducting boundary, where 
$\Delta_{\uparrow}=0$) transverse spin susceptibility to $\chi_{L}$ via mode
coupling effects. Because $\chi_{L}$ serves as the pairing potential for
the $\Delta_{\uparrow}$ ordering, this suggest that in the superconducting
phase the pairing potential decreases, since $\chi_{T}$
decreases for all wavenumber and frequencies with increasing 
$\Delta_{\uparrow}$. For spin-triplet superconductivity in a magnetic field 
one generically expects an ordering of the spin down electrons, 
$\Delta_{\downarrow}\neq 0$, at some temperature 
$T_{\text{c}\downarrow} <  T_{\text{c}\uparrow}$. Since the 
pairing potential for $\Delta_{\downarrow}$ is also $\chi_{\text{L}}$, this 
implies there is a weaker tendency for the onset of $\Delta_{\downarrow}$ 
given $\Delta_{\uparrow}\neq 0$, than there is for the onset of 
$\Delta_{\uparrow}$ ordering. Another general mechanism decreasing
$T_{\text{c}\downarrow}$ compared to $T_{\text{c}\uparrow}$ is that the 
density of states at the Fermi surface for the down-spin electrons decreases 
with increasing magnetization. Naively, one therefore expects that for small
magnetization the cutoff in Eq.\ (\ref{eq:2}), which is proportional to 
$\delta^{-2}$, greatly suppresses
the $\Delta _{\downarrow }$ ordering, while for large $\delta$ the density
of state effect should decrease it. This would result in, for example, a
specific heat discontinuity at $T_{\text{c}\uparrow}$, and a substantial
residual specific heat coefficient at lower $T$, since 
the down-spin electron would not undergo a superconducting transition until 
perhaps immeasurably low temperatures. This is certainly consistent with 
the experimental results in the URhGe system, and it is the interpretation 
given in Ref.\ \onlinecite{Aoki_et_al_01}. However, our detailed numerical 
work suggests that this is not the generic situation for the pairing 
mechanism we have considered. Clearly, with improving sample quality that 
will sharpen the feature in the specific heat, this question can be decided
experimentally.

We thank Meigan Aronson and Thomas Vojta for discussions.
This work was supported by the NSF grant under Nos. DMR-01-32555 and
DMR-01-32726.


\begin{thebibliography}{27}
\expandafter\ifx\csname natexlab\endcsname\relax\def\natexlab#1{#1}\fi
\expandafter\ifx\csname bibnamefont\endcsname\relax
  \def\bibnamefont#1{#1}\fi
\expandafter\ifx\csname bibfnamefont\endcsname\relax
  \def\bibfnamefont#1{#1}\fi
\expandafter\ifx\csname citenamefont\endcsname\relax
  \def\citenamefont#1{#1}\fi
\expandafter\ifx\csname url\endcsname\relax
  \def\url#1{\texttt{#1}}\fi
\expandafter\ifx\csname urlprefix\endcsname\relax\def\urlprefix{URL }\fi
\providecommand{\bibinfo}[2]{#2}
\providecommand{\eprint}[2][]{\url{#2}}

\bibitem[{\citenamefont{Saxena et~al.}(2000)\citenamefont{Saxena, Agarwal,
  Ahilan, Grosche, Haselwimmer, Steiner, Pugh, Walker, Julian, Monthoux
  et~al.}}]{Saxena_et_al_00}
\bibinfo{author}{\bibfnamefont{S.~S.} \bibnamefont{Saxena}},
  \bibinfo{author}{\bibfnamefont{P.}~\bibnamefont{Agarwal}},
  \bibinfo{author}{\bibfnamefont{K.}~\bibnamefont{Ahilan}},
  \bibinfo{author}{\bibfnamefont{F.~M.} \bibnamefont{Grosche}},
  \bibinfo{author}{\bibfnamefont{R.~K.~W.} \bibnamefont{Haselwimmer}},
  \bibinfo{author}{\bibfnamefont{M.~J.} \bibnamefont{Steiner}},
  \bibinfo{author}{\bibfnamefont{E.}~\bibnamefont{Pugh}},
  \bibinfo{author}{\bibfnamefont{I.~R.} \bibnamefont{Walker}},
  \bibinfo{author}{\bibfnamefont{S.~R.} \bibnamefont{Julian}},
  \bibinfo{author}{\bibfnamefont{P.}~\bibnamefont{Monthoux}},
  \bibnamefont{et~al.}, \bibinfo{journal}{Nature}
  \textbf{\bibinfo{volume}{406}}, \bibinfo{pages}{587} (\bibinfo{year}{2000}).

\bibitem[{\citenamefont{Pfleiderer et~al.}(2001)\citenamefont{Pfleiderer,
  Uhlarz, Hayden, Vollmer, von L{\"o}hneysen, Bernhoeft, and
  Lonzarich}}]{Pfleiderer_et_al_01}
\bibinfo{author}{\bibfnamefont{C.}~\bibnamefont{Pfleiderer}},
  \bibinfo{author}{\bibfnamefont{M.}~\bibnamefont{Uhlarz}},
  \bibinfo{author}{\bibfnamefont{S.~M.} \bibnamefont{Hayden}},
  \bibinfo{author}{\bibfnamefont{R.}~\bibnamefont{Vollmer}},
  \bibinfo{author}{\bibfnamefont{H.}~\bibnamefont{von L{\"o}hneysen}},
  \bibinfo{author}{\bibfnamefont{N.~R.} \bibnamefont{Bernhoeft}},
  \bibnamefont{and} \bibinfo{author}{\bibfnamefont{G.~G.}
  \bibnamefont{Lonzarich}}, \bibinfo{journal}{Nature}
  \textbf{\bibinfo{volume}{412}}, \bibinfo{pages}{58} (\bibinfo{year}{2001}).

\bibitem[{\citenamefont{Aoki et~al.}(2001)\citenamefont{Aoki, Huxley,
  Ressouche, Braithwaite, Floquet, Brison, Lhotel, and
  Paulsen}}]{Aoki_et_al_01}
\bibinfo{author}{\bibfnamefont{D.}~\bibnamefont{Aoki}},
  \bibinfo{author}{\bibfnamefont{A.}~\bibnamefont{Huxley}},
  \bibinfo{author}{\bibfnamefont{E.}~\bibnamefont{Ressouche}},
  \bibinfo{author}{\bibfnamefont{D.}~\bibnamefont{Braithwaite}},
  \bibinfo{author}{\bibfnamefont{J.}~\bibnamefont{Floquet}},
  \bibinfo{author}{\bibfnamefont{J.~P.} \bibnamefont{Brison}},
  \bibinfo{author}{\bibfnamefont{E.}~\bibnamefont{Lhotel}}, \bibnamefont{and}
  \bibinfo{author}{\bibfnamefont{C.}~\bibnamefont{Paulsen}},
  \bibinfo{journal}{Nature} \textbf{\bibinfo{volume}{413}},
  \bibinfo{pages}{613} (\bibinfo{year}{2001}).

\bibitem[{\citenamefont{Fay and Appel}(1980)}]{Fay_Appel_80}
\bibinfo{author}{\bibfnamefont{D.}~\bibnamefont{Fay}} \bibnamefont{and}
  \bibinfo{author}{\bibfnamefont{J.}~\bibnamefont{Appel}},
  \bibinfo{journal}{Phys. Rev. B} \textbf{\bibinfo{volume}{22}},
  \bibinfo{pages}{3173} (\bibinfo{year}{1980}).

\bibitem[{\citenamefont{Shick and Pickett}(2001)}]{Shick_Picket_01}
\bibinfo{author}{\bibfnamefont{A.}~\bibnamefont{Shick}} \bibnamefont{and}
  \bibinfo{author}{\bibfnamefont{W.}~\bibnamefont{Pickett}},
  \bibinfo{journal}{Phys. Rev. Lett.} \textbf{\bibinfo{volume}{86}},
  \bibinfo{pages}{300} (\bibinfo{year}{2001}).

\bibitem[{\citenamefont{Santi et~al.}(2001)\citenamefont{Santi, Dugdale, and
  Jarlborg}}]{Santi_et_al_01}
\bibinfo{author}{\bibfnamefont{G.}~\bibnamefont{Santi}},
  \bibinfo{author}{\bibfnamefont{S.~B.} \bibnamefont{Dugdale}},
  \bibnamefont{and} \bibinfo{author}{\bibfnamefont{T.}~\bibnamefont{Jarlborg}},
  \bibinfo{journal}{Phys. Rev. Lett.} \textbf{\bibinfo{volume}{87}},
  \bibinfo{pages}{247004} (\bibinfo{year}{2001}).

\bibitem[{\citenamefont{Machida and Ohmi}(2001)}]{Machida_Ohmi_01}
\bibinfo{author}{\bibfnamefont{K.}~\bibnamefont{Machida}} \bibnamefont{and}
  \bibinfo{author}{\bibfnamefont{T.}~\bibnamefont{Ohmi}},
  \bibinfo{journal}{Phys. Rev. Lett.} \textbf{\bibinfo{volume}{86}},
  \bibinfo{pages}{850} (\bibinfo{year}{2001}).

\bibitem[{\citenamefont{Shimahara and Kohmoto}(2002)}]{Shimahara_Kohmoto_02}
\bibinfo{author}{\bibfnamefont{H.}~\bibnamefont{Shimahara}} \bibnamefont{and}
  \bibinfo{author}{\bibfnamefont{M.}~\bibnamefont{Kohmoto}},
  \bibinfo{journal}{Europhys. Lett.} \textbf{\bibinfo{volume}{57}},
  \bibinfo{pages}{247} (\bibinfo{year}{2002}).

\bibitem[{\citenamefont{Watanabe and Miyake}(2002)}]{Watanabe_Miyake_02}
\bibinfo{author}{\bibfnamefont{S.}~\bibnamefont{Watanabe}} \bibnamefont{and}
  \bibinfo{author}{\bibfnamefont{K.}~\bibnamefont{Miyake}},
  \bibinfo{journal}{J. Phys. Soc. Jpn.} \textbf{\bibinfo{volume}{71}},
  \bibinfo{pages}{2489} (\bibinfo{year}{2002}).

\bibitem[{\citenamefont{Sandeman et~al.}(2003)\citenamefont{Sandeman,
  Lonzarich, and Schofield}}]{Sandeman_et_al_03}
\bibinfo{author}{\bibfnamefont{K.}~\bibnamefont{Sandeman}},
  \bibinfo{author}{\bibfnamefont{G.}~\bibnamefont{Lonzarich}},
  \bibnamefont{and}
  \bibinfo{author}{\bibfnamefont{A.}~\bibnamefont{Schofield}},
  \bibinfo{journal}{Phys. Rev. Lett.} \textbf{\bibinfo{volume}{90}},
  \bibinfo{pages}{167005} (\bibinfo{year}{2003}).

\bibitem[{\citenamefont{Kirkpatrick et~al.}(2001)\citenamefont{Kirkpatrick,
  Belitz, Vojta, and Narayanan}}]{us_p-wave_letter}
\bibinfo{author}{\bibfnamefont{T.~R.} \bibnamefont{Kirkpatrick}},
  \bibinfo{author}{\bibfnamefont{D.}~\bibnamefont{Belitz}},
  \bibinfo{author}{\bibfnamefont{T.}~\bibnamefont{Vojta}}, \bibnamefont{and}
  \bibinfo{author}{\bibfnamefont{R.}~\bibnamefont{Narayanan}},
  \bibinfo{journal}{Phys. Rev. Lett.} \textbf{\bibinfo{volume}{87}},
  \bibinfo{pages}{127003} (\bibinfo{year}{2001}).

\bibitem[{\citenamefont{Kirkpatrick and Belitz}(2003)}]{us_p-wave}
\bibinfo{author}{\bibfnamefont{T.~R.} \bibnamefont{Kirkpatrick}}
  \bibnamefont{and} \bibinfo{author}{\bibfnamefont{D.}~\bibnamefont{Belitz}},
  \bibinfo{journal}{Phys. Rev. B} \textbf{\bibinfo{volume}{67}},
  \bibinfo{pages}{024515} (\bibinfo{year}{2003}).

\bibitem[{\citenamefont{Belitz and Kirkpatrick}(2003)}]{us_unpublished_03}
\bibinfo{author}{\bibfnamefont{D.}~\bibnamefont{Belitz}} \bibnamefont{and}
  \bibinfo{author}{\bibfnamefont{T.~R.} \bibnamefont{Kirkpatrick}},
  \bibinfo{journal}{unpublished results}  (\bibinfo{year}{2003}).

\bibitem[{\citenamefont{Forster}(1975)}]{Forster_75}
\bibinfo{author}{\bibfnamefont{D.}~\bibnamefont{Forster}},
  \emph{\bibinfo{title}{Hydrodynamic Fluctuations, Broken Symmetry, and
  Correlation Functions}} (\bibinfo{publisher}{Benjamin, Reading, MA},
  \bibinfo{year}{1975}).

\bibitem[{Gol()}]{Goldstone_mode_footnote}
\bibinfo{note}{This is true if the superconducting order parameter is held
  fixed while rotating the magnetization. There is a Goldstone mode connected
  to the invariance under rotating {\em all} spins, which is unbroken.
  Estimates of the stiffness of this Goldstone mode, and its coupling to the
  spin density, show that the resulting contribution to the dimensionless
  $\chi_{\text{T}}$ is of $O(\mu^2/\tilde{{\bm k}}^2)$. Consequently, Eq.\
  (\ref{eq:2}) will hold for wavenumbers $\tilde{\bm k}^2 \agt \mu^4$, while
  for $\tilde{\bm k}^2 < \mu^4$, $\chi_{\text{T}}$ will diverge like $1/{\bm
  k}^2$ after all. Since wavenumbers on the order of $\mu$ are already hard to
  observe, this means that for practical purposes $\chi_{\text{T}}$ will appear
  massive.}

\bibitem[{\citenamefont{Aronson}(2003)}]{Aronson_03}
\bibinfo{author}{\bibfnamefont{M.}~\bibnamefont{Aronson}},
  \bibinfo{journal}{private communication}  (\bibinfo{year}{2003}).

\bibitem[{spe()}]{specific_heat_footnote}
\bibinfo{note}{This is the asymptotic low-$T$ behavior that results from the
  order parameter given in Eq.\ (\ref{eq:4}) below. For simplicity, we use the
  asymptotic behavior everywhere below $T_c$. The specific heat discontinuity
  has been chosen to match the maximum in the experimental data. Theoretical
  values depend on the symmetry of the order parameter. Typical values in a
  weak-coupling theory, which are universal, are on the order of 1.2 - 1.4
  (i.e., much larger than the height of the observed feature); in
  strong-coupling theories they are not universal
  \protect{\cite{Vollhardt_Woelfle_90}}.}

\bibitem[{orb()}]{orbital_structure_footnote}
\bibinfo{note}{This choice of the orbital structure of the gap function is
  entirely arbitrary. Determining theoretically which of the possible
  structures has the lowest energy would be very difficult. However, since
  different orbital symmetries lead to different thermodynamic properties, one
  can distinguish between them experimentally. For instance, an orbital
  symmetry corresponding to the ABM state in Helium 3 would lead to a specific
  heat $C(T\rightarrow 0)\propto T^3$ rather than $T^2$
  \protect{\cite{Vollhardt_Woelfle_90}}. For simplicity we only study the order
  parameter given in Eq.\ (\ref{eq:4}). If experimental evidence should favor a
  different structure, the theory can be easily adjusted to that.}

\bibitem[{\citenamefont{Vollhardt and W{\"o}lfle}(1990)}]{Vollhardt_Woelfle_90}
\bibinfo{author}{\bibfnamefont{D.}~\bibnamefont{Vollhardt}} \bibnamefont{and}
  \bibinfo{author}{\bibfnamefont{P.}~\bibnamefont{W{\"o}lfle}},
  \emph{\bibinfo{title}{The Superfluid Phases of Helium 3}}
  (\bibinfo{publisher}{Taylor \& Francis}, \bibinfo{year}{1990}).

\bibitem[{fin()}]{finite_gamma_footnote}
\bibinfo{note}{A ferromagnetic superconductor should have an intrinsic nonzero
  specific heat coefficient, in analogy to the specific heat in the vortex
  phase of an ordinary type-II superconductor
  \protect{\cite{Fetter_Hohenberg_69}}. Here we neglect this effect, whose size
  will depend on details of the superconducting phase which are not known.}

\bibitem[{\citenamefont{Fisher et~al.}(1989)\citenamefont{Fisher, Kim,
  Woodfield, Phillips, Taillefer, Hasselbach, Flouquet, Georgi, and
  Smith}}]{Fisher_et_al_89}
\bibinfo{author}{\bibfnamefont{R.~A.} \bibnamefont{Fisher}},
  \bibinfo{author}{\bibfnamefont{S.}~\bibnamefont{Kim}},
  \bibinfo{author}{\bibfnamefont{B.~F.} \bibnamefont{Woodfield}},
  \bibinfo{author}{\bibfnamefont{N.~E.} \bibnamefont{Phillips}},
  \bibinfo{author}{\bibfnamefont{L.}~\bibnamefont{Taillefer}},
  \bibinfo{author}{\bibfnamefont{K.}~\bibnamefont{Hasselbach}},
  \bibinfo{author}{\bibfnamefont{J.}~\bibnamefont{Flouquet}},
  \bibinfo{author}{\bibfnamefont{A.~L.} \bibnamefont{Georgi}},
  \bibnamefont{and} \bibinfo{author}{\bibfnamefont{J.~L.} \bibnamefont{Smith}},
  \bibinfo{journal}{Phys. Rev. Lett.} \textbf{\bibinfo{volume}{62}},
  \bibinfo{pages}{1411} (\bibinfo{year}{1989}).

\bibitem[{\citenamefont{Hasselbach et~al.}(1989)\citenamefont{Hasselbach,
  Taillefer, and Flouquet}}]{Hasselbach_et_al_89}
\bibinfo{author}{\bibfnamefont{K.}~\bibnamefont{Hasselbach}},
  \bibinfo{author}{\bibfnamefont{L.}~\bibnamefont{Taillefer}},
  \bibnamefont{and} \bibinfo{author}{\bibfnamefont{J.}~\bibnamefont{Flouquet}},
  \bibinfo{journal}{Phys. Rev. Lett.} \textbf{\bibinfo{volume}{63}},
  \bibinfo{pages}{93} (\bibinfo{year}{1989}).

\bibitem[{\citenamefont{Ma}(1976)}]{Ma_76}
\bibinfo{author}{\bibfnamefont{S.-K.} \bibnamefont{Ma}},
  \emph{\bibinfo{title}{Modern Theory of Critical Phenomena}}
  (\bibinfo{publisher}{Benjamin, Reading, MA}, \bibinfo{year}{1976}).

\bibitem[{RPA()}]{RPA_footnote}
\bibinfo{note}{In a diagrammatic language, this is equivalent to a generalized
  random-phase approximation, as is obvious from the structure of Eq.\
  (\ref{eq:8}).}

\bibitem[{\citenamefont{Belitz et~al.}(1998)\citenamefont{Belitz, Kirkpatrick,
  Millis, and Vojta}}]{us_magnon_dispersion}
\bibinfo{author}{\bibfnamefont{D.}~\bibnamefont{Belitz}},
  \bibinfo{author}{\bibfnamefont{T.~R.} \bibnamefont{Kirkpatrick}},
  \bibinfo{author}{\bibfnamefont{A.~J.} \bibnamefont{Millis}},
  \bibnamefont{and} \bibinfo{author}{\bibfnamefont{T.}~\bibnamefont{Vojta}},
  \bibinfo{journal}{Phys. Rev. B} \textbf{\bibinfo{volume}{58}},
  \bibinfo{pages}{14155} (\bibinfo{year}{1998}).

\bibitem[{\citenamefont{Br{\'e}zin and Wallace}(1973)}]{Brezin_Wallace_73}
\bibinfo{author}{\bibfnamefont{E.}~\bibnamefont{Br{\'e}zin}} \bibnamefont{and}
  \bibinfo{author}{\bibfnamefont{D.~J.} \bibnamefont{Wallace}},
  \bibinfo{journal}{Phys. Rev. B} \textbf{\bibinfo{volume}{7}},
  \bibinfo{pages}{1967} (\bibinfo{year}{1973}).

\bibitem[{\citenamefont{Fetter and Hohenberg}(1969)}]{Fetter_Hohenberg_69}
\bibinfo{author}{\bibfnamefont{A.~L.} \bibnamefont{Fetter}} \bibnamefont{and}
  \bibinfo{author}{\bibfnamefont{P.~C.} \bibnamefont{Hohenberg}}, in
  \emph{\bibinfo{booktitle}{Superconductivity}}, edited by
  \bibinfo{editor}{\bibfnamefont{R.~D.} \bibnamefont{Parks}}
  (\bibinfo{publisher}{Marcel Dekker, New York}, \bibinfo{year}{1969}), p.
  \bibinfo{pages}{817}.

\end{thebibliography}

\end{document}